\documentclass[aps,pra,showpacs]{revtex4}
\usepackage{eurosym}
\usepackage{graphicx}
\usepackage{amsmath}
\usepackage{amsfonts}
\usepackage{amssymb}
\usepackage[caption=false]{subfig}
\usepackage{float}

\begin{document}

\title{Dynamics of nonlinear-Schr\"{o}dinger breathers in a potential trap}
\author{B. A. Malomed$^{1,2,3},$ N. N. Rosanov$^{3,4,5}$, and S. V. Fedorov$%
^{3}$}
\affiliation{$^{1}$Department of Physical Electronics, School of Electrical Engineering,
Faculty of Engineering, Tel Aviv University, P.O.B. 39040, Ramat Aviv, Tel
Aviv, Israel\\
$^{2}$Center for Light-Matter Interaction, Tel Aviv University, P.O.B.
39040, Ramat Aviv, Tel Aviv, Israel\\
$^{3}$Saint Petersburg National Research University of Information
Technologies, Mechanics and Optics (ITMO University), 197101 Saint
Petersburg, Russia\\
$^{4}$Vavilov State Optical Institute, 199053 Saint Petersburg, Russia\\
$^{5}$Ioffe Physical-Technical Institute, Russian Academy of Sciences,
194021 Saint Petersburg, Russia}

\begin{abstract}
We consider the evolution of the 2-soliton (breather) of the nonlinear Schr%
\"{o}dinger equation on a semi-infinite line with the zero boundary
condition and a linear potential, which corresponds to the gravity field in
the presence of a hard floor. This setting can be implemented in atomic
Bose-Einstein condensates, and in a nonlinear planar waveguide in optics. In
the absence of the gravity, repulsion of the breather from the floor leads
to its splitting into constituent fundamental solitons, if the initial
distance from the floor is smaller than a critical value; otherwise, the
moving breather persists. In the presence of the gravity, the breather
always splits into a pair of \textquotedblleft co-hopping" fundamental
solitons, which may be frequency-locked in the form of a \textit{%
quasi-breather}, or unlocked, forming an incoherent \textit{pseudo-breather}%
. Some essential results are obtained in an analytical form, in addition to
the systematic numerical investigation.
\end{abstract}

\maketitle

\section{Introduction}

The nonlinear Schr\"{o}dinger (NLS) equation is a fundamental model for a
broad class of physical settings combining weak nonlinearity and weak linear
dispersion or diffraction \cite{book}. It is commonly known that, in the
absence of additional terms, the NLS equations with either self-focusing or
defocusing sign of the nonlinearity are integrable, the former one giving
rise to exact single- and multi-soliton solutions. An essential extension of
the concept of fundamental single-soliton states is provided by the class of
higher-order $n$-solitons ($n=2,3,4...$), which are produced, as exact
solutions, by the input in the form of the fundamental soliton multiplied by
integer $n$ \cite{Satsuma}. Alternatively, one can create a fundamental
soliton, corresponding to $n=1$, and apply a \textit{quench} of the
nonlinearity strength, making it stronger by a factor of $n^{2}$ \cite%
{quench1}-\cite{quench7}, which may be implemented, in particular, in atomic
Bose-Einstein condensates (BECs) by means of the Feshbach resonance \cite{FR}%
. Higher-order solitons are bound states of $n$ fundamental ones with
unequal amplitudes, whose binding energy is exactly zero (therefore, they
are subject to weak splitting instability). They perform periodic
oscillations (at a frequency which does not depend on $n$), hence the name
of \textquotedblleft breathers". In particular, the $2$- and $3$-solitons
are built as bound states of fundamental solitons with ratios of amplitudes
and norms $3:1$ and $5:3:1$, respectively \cite{Satsuma}.

In spite of the above-mentioned weak splitting instability, $n$-solitons,
and, first of all, $2$-solitons are relevant self-trapped modes, as they are
readily generated experimentally, along with fundamental solitons and other
varieties of multi-soliton states, in nonlinear optics \cite{Mollenauer}-%
\cite{KA}, as well as in BEC \cite{Canberra}, magnetic media \cite{magnetic}%
, superconductors \cite{super}, and in other settings. In particular, a
relevant issue is interaction of breathers with local defects and walls, as
well as dynamics of breathers trapped in potential wells. The latter is the
subject of the present work, as concerns the $2$-solitons. The interaction
with defects (including nonlinear potential barriers or traps, represented
by narrow regions carrying strong nonlinearity \cite{Bulgaria2}-\cite{NJP}),
walls, and potential wells was studied in detail for fundamental bright
solitons \cite{RMP}-\cite{Genoud}, but not for the breathers.

In this work, we address the dynamics of $2$-solitons in the framework of
the NLS equation for wave function $\psi (z,t)$ on a semi-infinite axis, $%
z\geq 0$, with the zero (reflective) boundary condition (b.c.), $\psi (z=0)=0
$. The equation, written in the scaled form, includes a linear potential $Gz$%
, which represents a constant force:

\begin{equation}
i\psi _{t}=-(1/2)\psi _{zz}-|\psi |^{2}\psi +Gz\psi ,  \label{GP}
\end{equation}%
the respective Hamiltonian being%
\begin{equation}
H=\int_{-\infty }^{+\infty }\left( \frac{1}{2}\left\vert \psi
_{z}\right\vert ^{2}dz-\frac{1}{2}\left\vert \psi \right\vert
^{4}dz+Gz\left\vert \psi \right\vert ^{2}\right) dz.  \label{E}
\end{equation}%
In addition to the Hamiltonian, Eq. (\ref{GP}) conserves the total norm,%
\begin{equation}
N=\int_{-\infty }^{+\infty }\left\vert u(z)\right\vert ^{2}dz.  \label{Norm}
\end{equation}

This model directly applies to BEC in a vertically or obliquely placed
cigar-shaped (quasi-one-dimensional) trap, under the action of gravity, with
the \textquotedblleft hard floor" at $z=0$ provided by a repelling laser
sheet \cite{gravity1,gravity2}. With temporal and spatial scales typical to
BEC experiments, $t_{0}\sim 1$ ms and $z_{0}\sim 1~\mathrm{\mu }$m, the
value of $G$ corresponding to the vertically placed trap holding atoms of $%
^{7}$Li under the action of the natural gravity field is $G\sim 1$. Below,
we consider essentially smaller values, \textit{viz}., $G=0.001,$ $0.01$,
and $0.1$, which correspond to oblique placement of the
quasi-one-dimensional trap, under small angles with the horizontal
direction, $\theta \sim 0.05^{\mathrm{o}}$, $0.5^{\mathrm{o}}$, and $5^{%
\mathrm{o}}$, respectively. Alternatively, this setting may be realized
under the action of microgravity \cite{micro}. Equation (\ref{GP}) with $t$
replaced by the propagation distance models the transmission of optical or
terahertz waves in planar nonlinear waveguides, with an edge at $z=0$, the
gravity term representing spatial modulation of the refractive index \cite%
{KA,Boardman}. In this case, the gradient of the refractive index,
corresponding to Eq. (\ref{GP}), is $\sim \left( G/100\right) \lambda ^{-1}$
in physical units, where $\lambda $ is the carrier wavelength. While this
value may be unrealistically high for $G\sim 1$ and a typical optical
wavelength, $\lambda \sim 1$ $\mathrm{\mu }$m, the estimate yields
reasonable values for terahertz radiation.

The rest of the paper is organized as follows. In Section II, we report some
analytical results which predict characteristic features of the $2$%
-soliton's dynamics in the present model. Results of systematic numerical
simulations, produced by means of the standard split-step Fourier-transform
algorithm, are reported in Section III, and the paper is concluded by
Section IV.

\section{Analytical estimates}

In the absence of b.c. $\psi (z=0)=0$ and linear potential $Gz$, the initial
condition $\psi \left( z,t=0\right) =2\eta ~\mathrm{sech}(\eta z)e^{iVz}$,
with arbitrary real constants $\eta $ and $V$, gives rise to the exact
breather (2-soliton) solution of integrable equation (\ref{GP}), oscillating
with frequency%
\begin{equation}
\omega _{\mathrm{br}}=4\eta ^{2}  \label{omega}
\end{equation}%
and moving with velocity $V$ \cite{Satsuma}:
\begin{gather}
\psi _{\mathrm{br}}\left( z,t\right) =4\eta \frac{\cosh \left( 3\eta \left(
z-Vt\right) \right) +3e^{4i\eta ^{2}t}\cosh \left( \eta \left( z-Vt\right)
\right) }{\cosh \left( 4\eta \left( z-Vt\right) \right) +4\cosh \left( 2\eta
\left( z-Vt\right) \right) +3\cos \left( 4\eta ^{2}t\right) }  \notag \\
\times \exp \left[ iVz+(i/2)\left( \eta ^{2}-V^{2}\right) t\right] .
\label{2}
\end{gather}%
Note that, while the breather periodically returns to the initial
configuration, with the single central maximum of density $\left\vert \psi
\left( z,t\right) \right\vert ^{2}$, at times $t=2\left( \pi /\omega _{%
\mathrm{br}}\right) m$, with integer $m$, at other times, $t=\left( \pi
/\omega _{\mathrm{br}}\right) (1+2m)$, the density profile features small
side maxima, separated from the central peak by distance%
\begin{equation}
\Delta z\approx \pm 1.32/\eta .  \label{Delta}
\end{equation}

If the b.c. \cite{Fokas,Manakov}, or the linear potential \cite{1976} are
separately added to the NLS equation, these terms do not break its
integrability. However, if combined together, they make Eq. (\ref{E}) a
\emph{nonintegrable} model of the potential trap, corresponding to the
effective potential%
\begin{equation}
U(z)=\left\{
\begin{array}{c}
Gz,~\mathrm{at~~}z>0, \\
\infty ,~\mathrm{at~~}z<0.%
\end{array}%
\right.  \label{trap}
\end{equation}

In terms of the inverse-scattering transform, exact solution (\ref{2}) is a
nonlinear superposition of two fundamental solitons which, in isolation,
have the form of
\begin{equation}
\psi _{1,2}=\eta _{1,2}\mathrm{sech}(\eta _{1,2}z)\exp \left[ (i/2)\left(
\eta _{1,2}^{2}-V^{2}\right) +iVz\right] ,  \label{sol}
\end{equation}%
whose amplitudes,
\begin{equation}
\eta _{1}=3\eta ,~\eta _{2}=\eta ,  \label{3:1}
\end{equation}%
are subject to the above-mentioned ratio, $\eta _{1}:\eta _{2}=3:1$, the
summary norm of these solitons being equal to the total norm of breather (%
\ref{2}): $2\left( \eta _{1}+\eta _{2}\right) =8\eta $ \cite{Satsuma}.
Because the binding energy of the $2$-soliton in the integrable equation is
exactly zero, it may fission into a pair of the constituent fundamental
solitons (\ref{sol}). In particular, weak time-periodic modulation of the
nonlinearity strength with frequency $\omega $ gives rise to resonant
fission at $\omega =\omega _{\mathrm{br}}$ \cite{HS}.

The first dynamical situation addressed by means of numerical simulations
below, in the absence of gravity ($G=0$), is to place the center of the $2$%
-soliton with zero velocity at point $z=z_{0}$, which corresponds to the
initial condition%
\begin{equation}
\psi \left( z,t=0\right) =2\eta ~\mathrm{sech}(\eta \left( z-z_{0}\right) ),~%
\mathrm{at}~~z>0.  \label{input}
\end{equation}%
The b.c. $\psi (z=0)=0$ suggests to replace this input by one combining the
actual input with its mirror image placed, with the opposite sign, at $z<0$
(so that the zero b.c. identically holds), thus considering Eq. (\ref{GP})
(with $G=0$, for the time being) on the infinite axis, with the extended
initial condition,%
\begin{equation}
\psi (z,t=0)=A~\left[ \mathrm{sech}\left( \eta \left( z-z_{0}\right) \right)
-\mathrm{sech}\left( \eta \left( z+z_{0}\right) \right) \right] .
\label{input-2}
\end{equation}%
where the common amplitude, $A$, is taken as a free parameter, to develop
the analysis in a more general form.

The repulsive interaction of the actual quasi-soliton input with its
negative mirror image pushes the quasi-soliton towards $z\rightarrow \infty $%
, our objective being to predict velocity $V$ which it will thus acquire. To
this end, following Ref. \cite{interaction} (see also Refs. \cite{Karpman}
and \cite{RMP}), we use an effective potential energy of the repulsive
interaction, which can be easily found for input (\ref{input-2}):
\begin{equation}
U_{\mathrm{int}}=8A^{2}\eta \exp \left( -2\eta z_{0}\right) ,  \label{U}
\end{equation}%
assuming $\eta z_{0}\gg 1$, cf. Ref. \cite{JETP}. Further, the total kinetic
energy of the quasi-soliton, moving with velocity $V$, and its mirror image
moving with velocity $-V$, is $K=NV^{2}$, taking into account the well-known
fact that the effective mass of each term in expression (\ref{input-2}) is
equal to its norm \cite{RMP} [see Eq. (\ref{Norm}), where the integration is
performed separately for each soliton]: $M_{\mathrm{eff}}=2A^{2}\eta $.
Lastly, the velocity of the established regime of motion is determined by
the energy-balance condition, $U_{\mathrm{int}}=K$, i.e.,%
\begin{equation}
V=2\exp \left( -\eta z_{0}\right) .  \label{Vel}
\end{equation}%
It is worthy to note that the result given by Eq. (\ref{Vel}) is general, in
the sense that amplitude $A$ cancels out in it, hence the predicted velocity
does not depend on the choice of the initial amplitude, while it depends on
the width, $\eta ^{-1}$. Furthermore, this dependence suggests that the
repulsion from the mirror image can make the breather unstable against the
fission into the constituent fundamental solitons, characterized by
different values $\eta _{1,2}$ [see Eq. (\ref{3:1})], as they give rise to
different velocities as per Eq. (\ref{Vel}).

Another analytical prediction, relevant for the comparison with numerical
results reported below, pertains to the effective equilibrium position of
quasi-soliton (\ref{input}) in the presence of the gravity, $G>0$. Indeed,
in this case, the last term in Hamiltonian (\ref{E}) gives rise to the
gravity energy, which we take as the sum of the respective terms for the
actual quasi-soliton and its mirror image. Combining it with interaction
energy (\ref{U}), we derive the total potential energy:
\begin{equation}
U_{\mathrm{tot}}(z_{0})=8A^{2}\eta \exp \left( -2\eta z_{0}\right)
+2GA^{2}\eta z_{0},  \label{U(z)}
\end{equation}%
which translates the underlying trapping energy (\ref{trap}) into the
effective potential for the soliton's central coordinate. The equilibrium
position coincides with the minimum of energy (\ref{U(z)}), $dU_{\mathrm{tot}%
}(z)/dz=0$, i.e.,%
\begin{equation}
z_{\mathrm{equil}}(G)=\left( 2\eta \right) ^{-1}\ln \left( 8\eta /G\right) ,
\label{zeq}
\end{equation}%
as recently demonstrated in Ref. \cite{JETP}.

\section{Numerical results}

\subsection{Zero-gravity case}

As mentioned above, simulations of Eq. (\ref{GP}) with b.c. $\psi (z=0)=0$
and input (\ref{input}) were first run without the gravity, $G=0$, to
identify effects of the interaction of the initial breather with its mirror
image. The results are displayed here for $\eta =\allowbreak 1.\,\allowbreak
504$, which is chosen for convenience of the presentation (the scaling
invariance of the model with $G=0$ makes all the values of $\eta $ mutually
equivalent).

\begin{figure}[tbp]
\centering
\includegraphics[width=0.8\textwidth]{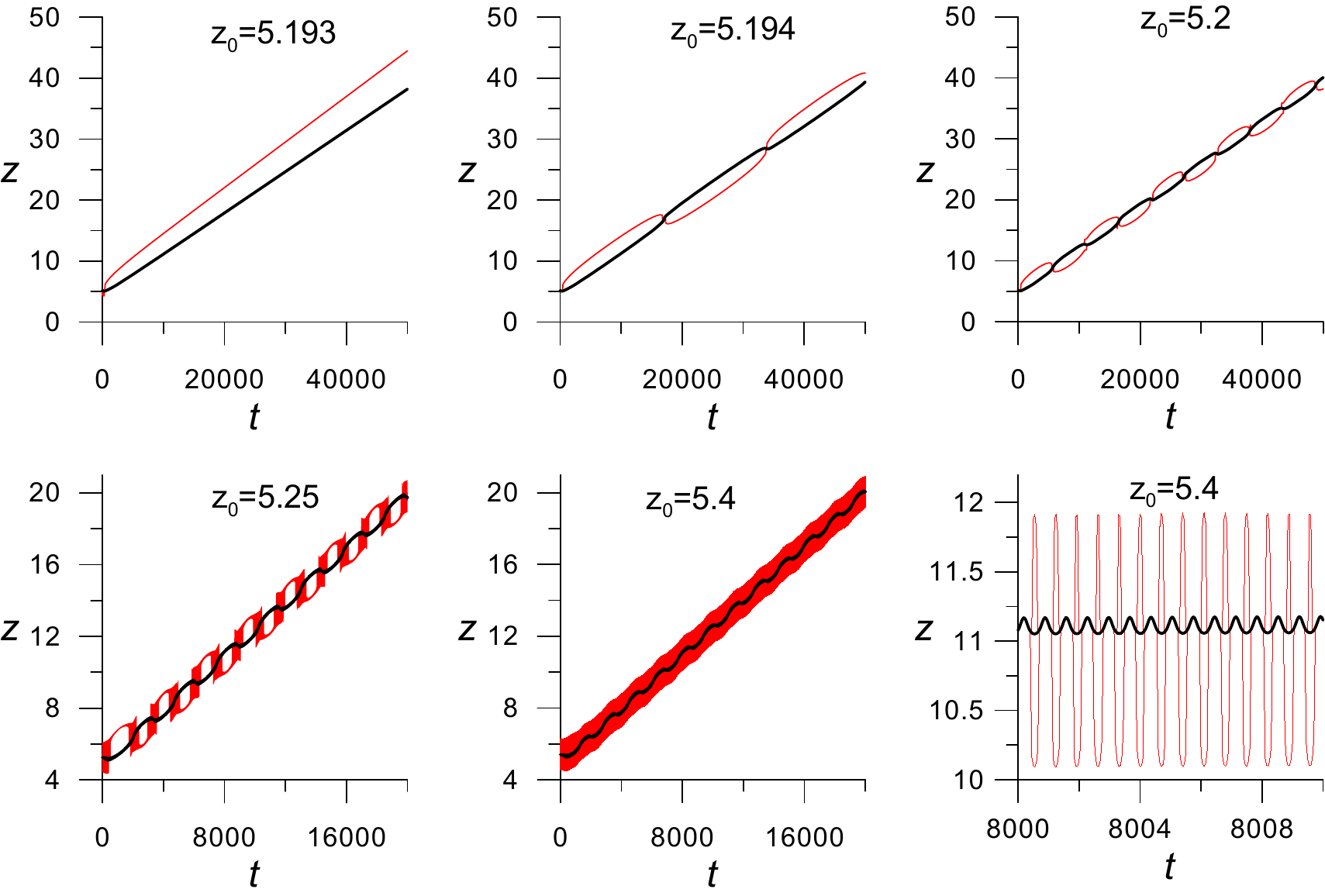} 
\caption{The motion of local maxima of density $\left\vert \protect\psi %
(z,t)\right\vert ^{2}$, as produced by numerical solutions of Eq. (\protect
\ref{GP}) in the absence of gravity ($G=0$) with input (\protect\ref{input})
taken at different values of $z_{0}$, which are indicated in individual
panels (the last panel displays details of the high-frequency oscillations
for $z_{0}=5.4$). Black and red lines depict, severally, the higher central
maximum, and lower side ones.}
\label{fig1}
\end{figure}

\begin{figure}[tbp]
\centering\subfloat[]{
\includegraphics[width=0.40\textwidth]{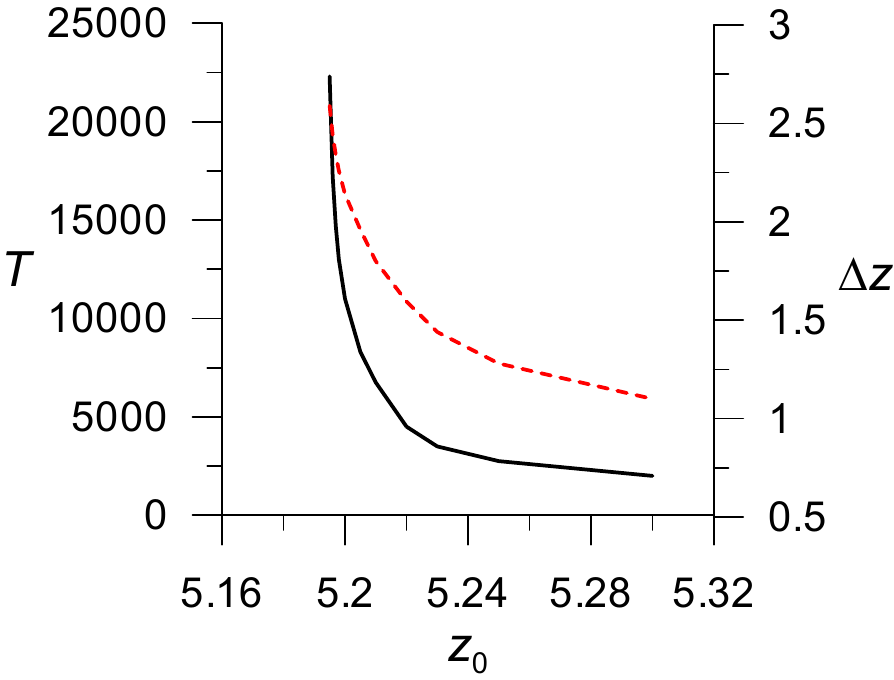}}
\subfloat[]{
\includegraphics[width=0.40\textwidth]{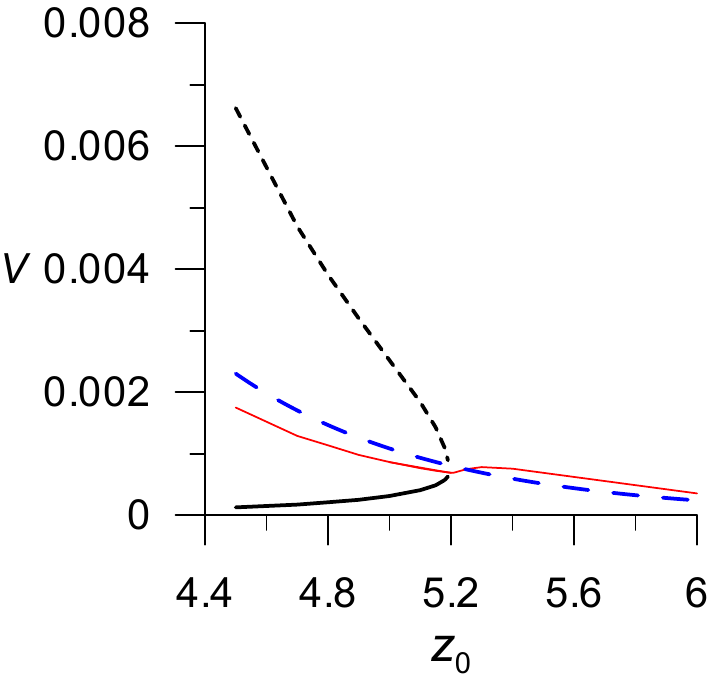}}
\caption{(a) The solid black and dashed red lines show, severally, period $T$
of oscillations of the moving bound state, and the respective largest
distance $\Delta z$ between the higher density maximum and the smaller one
oscillating around it, see Fig. \protect\ref{fig1}, vs. the initial
position, $z_{0}$, in the absence of gravity, $G=0$. (b) Solid and dashed
black lines show, respectively, velocities $V$ of motion of the heavy and
light fundamental solitons, in the case of the fission of the 2-soliton,
i.e., at $z<5.193$. The red line is the velocity of the motion of the center
of mass, for the split and unsplit states alike. The dashed blue line shows
the analytical prediction given by Eq. (\protect\ref{Vel}) for the velocity
of the unsplit $2$-soliton, or for the center-of-mass velocity of the pair
of separating solitons. }
\label{fig2}
\end{figure}

An essential finding is the existence of a critical value of the initial
position of the $2$-soliton's center,
\begin{equation}
\left( z_{0}\right) _{\mathrm{cr}}\approx 5.194,  \label{cr}
\end{equation}
such that, as conjectured above on the basis of Eq. (\ref{Vel}), the $2$%
-soliton, originally placed with zero velocity at $z_{0}<\left( z_{0}\right)
_{\mathrm{cr}}$, splits into two fundamental solitons, precisely with the
expected amplitudes predicted by Eq. (\ref{3:1}), as shown in the panel of
Fig. \ref{fig1} pertaining to $z_{0}=5.193$. The released fundamental
solitons slowly separate, moving towards $z\rightarrow \infty $. In the case
of $5.194\leq z_{0}<5.4$, the breather survives in the form of a moving
oscillatory bound state, as in this case the attraction between the
constituent fundamental solitons is sufficient to overcome the splitting
factor, which attenuates exponentially with the increase of $z_{0}$, as per
Eq. (\ref{Vel}). The velocity of the moving bound state is accurately
predicted by Eq. (\ref{Vel}) [see Fig. \ref{fig2}(b) below], while its
oscillation frequency is much lower than the standard value given by Eq. (%
\ref{omega}), as seen in panels of Fig. \ref{fig1} pertaining to $z_{0}=5.194
$, $5.2$, and $5.25$. Eventually, at $z_{0}\geq 5.4$, the moving breather
restores the standard oscillation frequency (\ref{omega}), as shown in the
panels of \ref{fig1} corresponding to $z=5.4$.

Note that, in the interval of $5.194\leq z_{0}<5.4$, the breather is
different from exact solution (\ref{2}), as it exhibits a single side
maximum oscillating around the central peak, unlike the pair of symmetric
side maxima in solution (\ref{2}), whose positions are given by Eq. (\ref%
{Delta}). Such generalized spatially asymmetric solutions for NLS breathers
are known too \cite{Gordon}. The usual symmetric shape is nearly restored at
$z_{0}\geq 5.4$. In particular, the largest distance between the side maxima
in the panel of Fig. \ref{fig1} pertaining to $z_{0}=5.4$ coincides with the
double value given by Eq. (\ref{Delta}).

Systematically collected results of the simulations are summarized in Fig. %
\ref{fig2}. Naturally, both the period of oscillations, $T$, and the
oscillation amplitude, $\Delta z$, diverge as the breather is approaching
the fission point, $z_{0}-5.193\rightarrow +0$. As concerns the velocity, it
is worthy to mention that the analytical prediction given by Eq. (\ref{Vel})
is quite accurate both for the unsplit breather and for the center of mass
of the split pair of the fundamental solitons, see the blue dashed curve in
panel \ref{fig2}(b).

Lastly, the above-mentioned scaling invariance of Eq. (\ref{GP}) with $G=0$
implies that, for other values of amplitude $\eta $ of input (\ref{input})
(recall the above results are displayed for fixed $\eta =1.504$), the
critical input's coordinate, $\left( z_{0}\right) _{\mathrm{cr}}$, can be
obtained from the above value (\ref{cr}), multiplying it by $\left(
1.504/\eta \right) ^{2}$.

\subsection{Formation of quasi- and pseudo-breathers in the presence of
gravity}

\begin{figure}[tbp]
\centering
\includegraphics[width=0.80\textwidth]{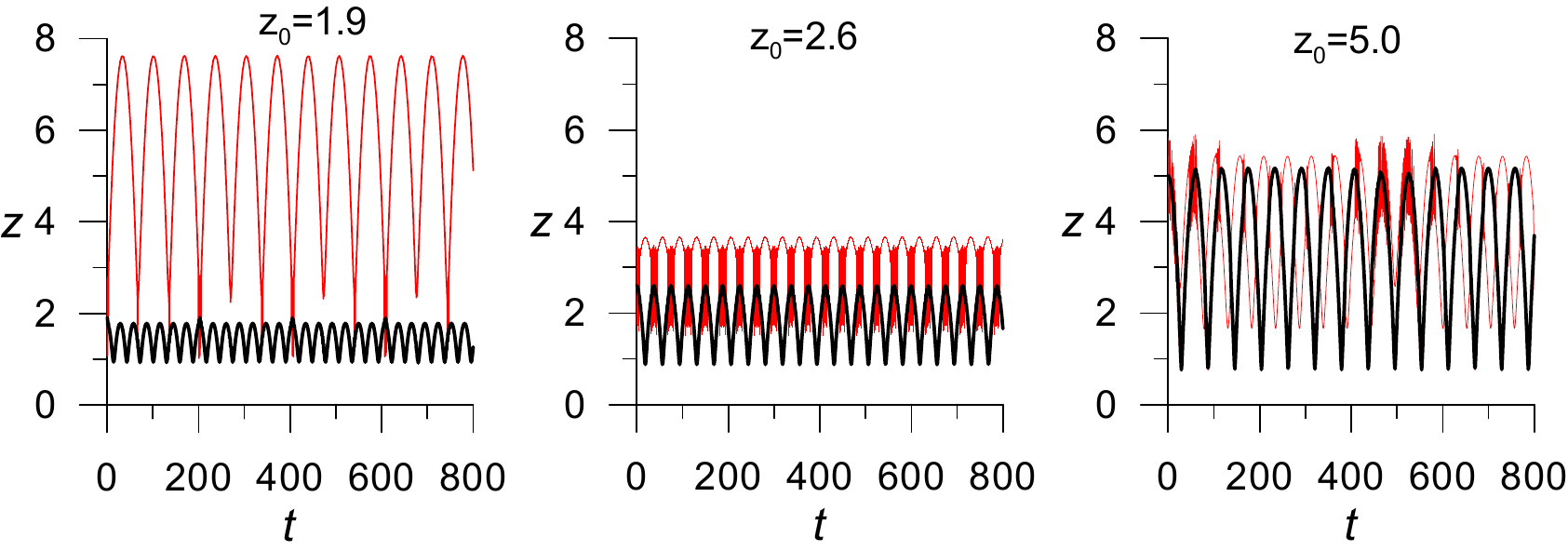} 
\caption{Black and red lines depict the motion of heavy and light
fundamental solitons, as produced by the fission of the $2$-soliton input (%
\protect\ref{input}) with values of $z_{0}$ indicated in the panels, in the
presence of the gravity with strength $G=0.01$.}
\label{fig3}
\end{figure}

\begin{figure}[tbp]
\centering
\includegraphics[width=0.80\textwidth]{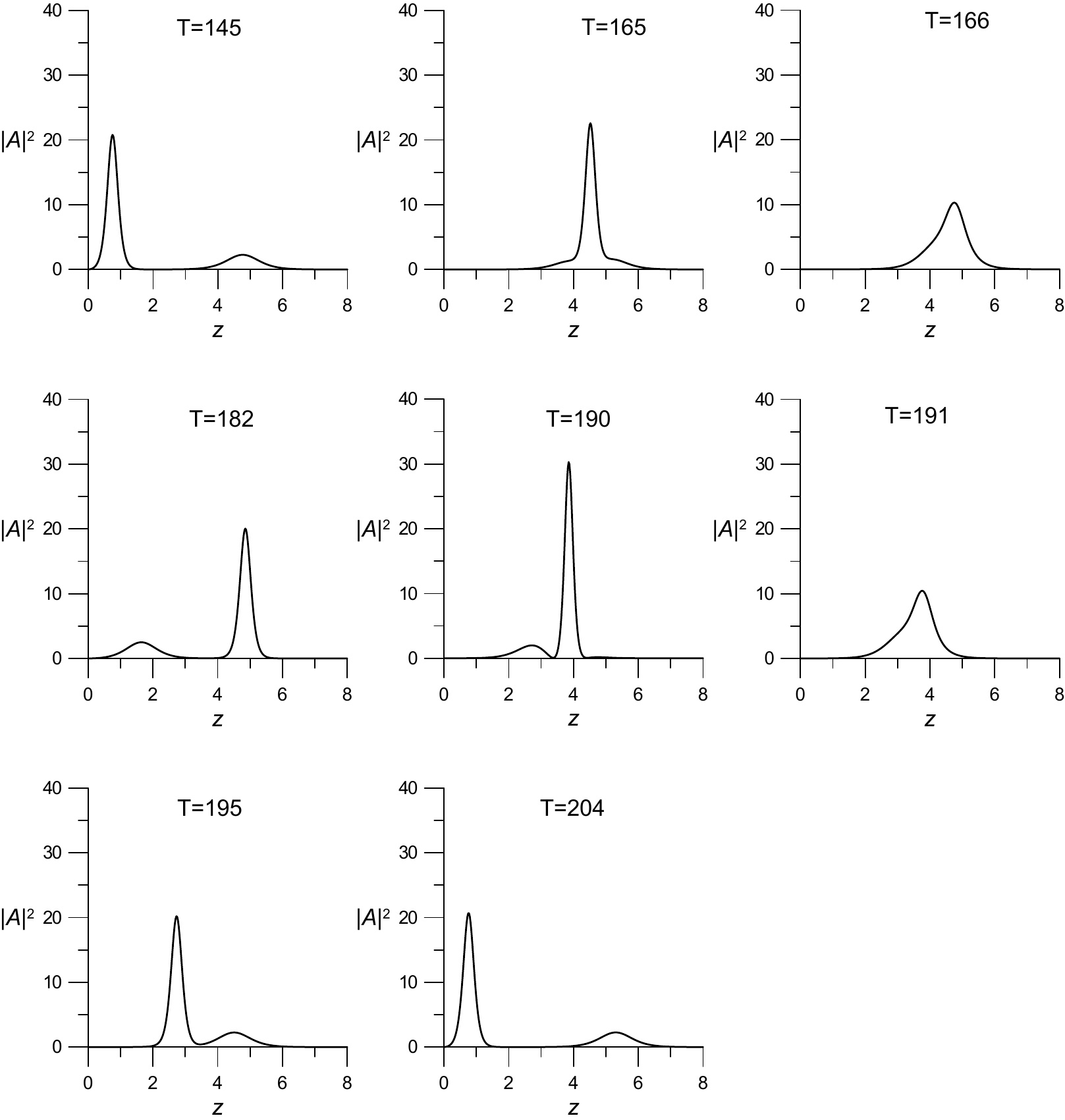} 
\caption{A set of snapshots corresponding to values of time ($\mathrm{T}$)
indicated in panels, which illustrate the dynamical regime of the
\textquotedblleft pseudo-breather" type, observed at $G=0.01$ and $z_{0}=5$.}
\label{fig4}
\end{figure}

If the gravity field is included in Eq. (\ref{GP}), the original breather
always suffers fission, but the gravity does not allow the splinters
(fundamental constituent solitons), or the breather as a whole, to escape,
unlike the case of $G=0$ considered above. As a result, the system relaxes
into a dynamical state which may be classified either as a \textquotedblleft
quasi-breather", originating from the input with $z_{0}$ relatively small,
or as a \textquotedblleft pseudo-breather", for larger $z_{0}$. In the
former case, the fundamental solitons periodically collide with relatively
small velocities, which makes the effective interaction between them strong
enough to frequency-lock their oscillatory (\textquotedblleft co-hopping")
motion, as seen in panels pertaining to $z_{0}=1.9$ and $2.6$ in Fig. \ref%
{fig3}, with the ratios of the locked frequencies of the heavy and light
solitons being, respectively, $3:1$ and $1:1$. On the other hand, the
dynamical regime originating from larger values of $z_{0}$, such as $%
z_{0}=5.0$ in Fig. \ref{fig3}, leads to collisions at relatively large
velocities, which attenuates the effective interaction, and prevents the
establishment of the frequency-locked regime. Instead, a quasi-random
pseudo-breather dynamical state is observed in the latter case,
\textquotedblleft pseudo" implying the absence of coherence between the
motion of the two fundamental solitons. A sequence of elastic collisions
between the heavy and light solitons in the latter case is illustrated by
Fig. \ref{fig4} (individual collisions look similar in the frequency-locked
quasi-breather regime).{\LARGE \ }

\begin{figure}[tbp]
\centering\subfloat[]{
\includegraphics[width=0.80\textwidth]{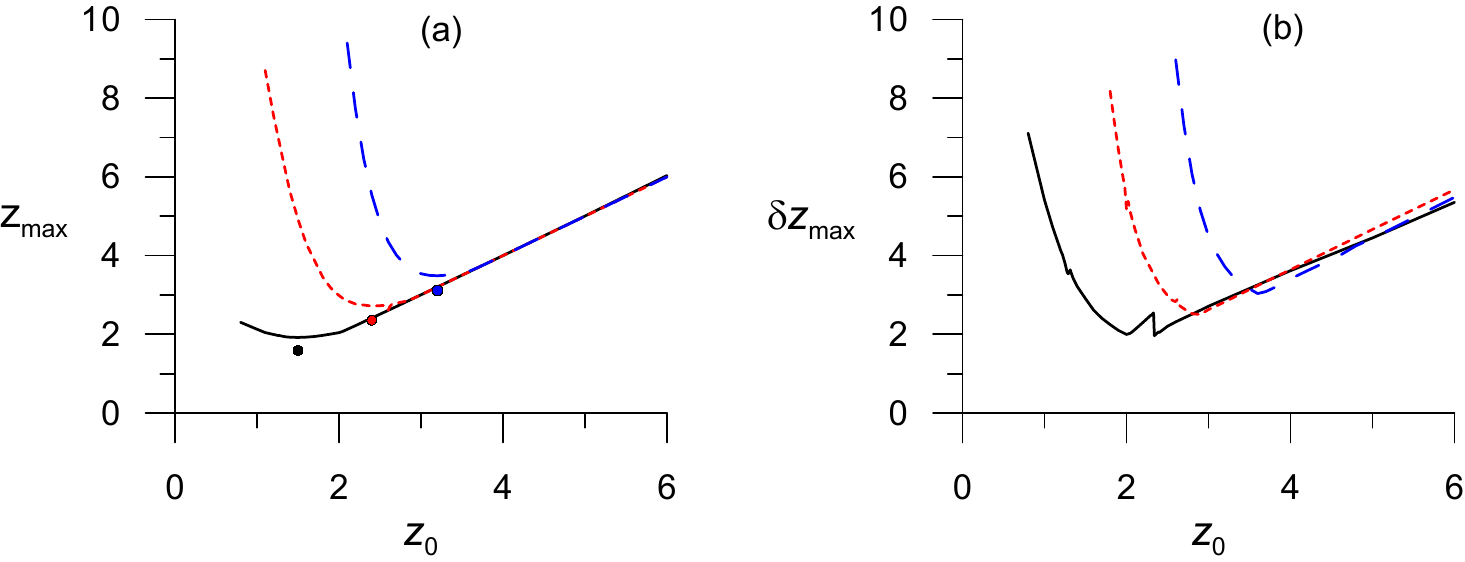}} 
\caption{(a) The largest separation between the center of mass of the two
``co-hopping" solitons and the ``hard floor" ($z=0$) vs. the initial
center-of-mass coordinate, $z_{0}$ [see Eq. (\protect\ref{input})]. (b) The
largest separation between the two solitons, $\protect\delta z_{\max }$, vs.
$z_{0}$. In both panels, black, red, and blue curves pertain to gravity
strengths $G=0.1$, $0.01$, and $0.001$, respectively. In (a), dots of the
same colors depict respective positions of the analytically predicted
potential minima given by Eq. (\protect\ref{zeq}).}
\label{fig5}
\end{figure}

Finally, characteristics of the ``co-hopping" regimes, namely, the largest
distance of the center-of-mass coordinate of the set of two solitons, $%
z_{\max }$, from the system's edge ($z=0$), and the largest separation
between the solitons, $\delta z_{\max }$, are summarized in Fig. \ref{fig5}.
In particular, minima of dependences $z_{\max }(z_{0})$ can be readily
predicted as the location of the potential minimum, $z_{\mathrm{equil}}$,
given by Eq. (\ref{zeq}) for the same $\eta $ as in unsplit input (\ref%
{input}). Indeed, for $z_{0}=$ $z_{\mathrm{equil}}$ the fission of the input
is expected to be minimal, while for $z_{0}\neq z_{\mathrm{equil}}$ strong
fission gives rise to large-amplitude oscillations of the constituent
solitons, leading to larger values of $z_{\max }$. Thus, for the values of $%
G $, which are represented in Fig. \ref{fig5}, and $\eta =1.504$ adopted
here, Eq. (\ref{zeq}) yields $z_{\mathrm{equil}}(G=0.1)\approx 1.59 $, $z_{%
\mathrm{equil}}(G=0.01)\approx 2.36$, $z_{\mathrm{equil}}(G=0.001)\approx
3.12$. These values are shown by dots with coordinates $z_{0}=z_{\max }=z_{%
\mathrm{equil}}(G)$ in Fig. \ref{fig5}(a), being, indeed, quite close to
values of $z_{0}$ at which the three curves feature their minima.

Lastly, small ``notches" observed on the curves in Fig. \ref{fig5}(b) may be
explained by effects of the emission of radiation from the moving solitons.
Detailed analysis of these weak features is beyond the scope of the present
paper.

\section{Conclusion}

Along with fundamental solitons, breathers, i.e., $2$-solitons, have drawn
much interest as collective excitations in diverse physical settings modeled
by the NLS equation. Here, we have addressed dynamics of breathers in the
semi-infinite system with a reflecting edge (\textquotedblleft hard floor")
and the linear potential, which represents gravity or a similar effective
field, in atomic BECs and nonlinear optical waveguides. In the absence of
the gravity, the repulsion of the $2$-soliton from its mirror image causes
its splitting in two constituent fundamental solitons, with the ratio of
norms and amplitudes $3:1$, if the initial distance of the $2$-soliton from
the edge is smaller than a critical value; otherwise, it moves away from the
edge in the form of a persistent breather. Inclusion of the gravity always
causes fission of the breather into the same pair of fundamental solitons.
They feature the \textquotedblleft co-hopping" motion, in which they may be
frequency-locked into a coherent quasi-breather, or remain in the form of an
incoherent pseudo-breather. In addition to the systematic numerical
simulations, basic characteristics of the considered dynamical regimes were
predicted analytically, with the help of the effective potential for the $2$%
-soliton input.

As an extension of the present analysis, it may be interesting to develop it
for $3$-solitons. A challenging issue is a possibility of the consideration
of multiple solitons in the quantum NLS model, cf. Refs. \cite{Heidelberg}-%
\cite{Drummond2}, \cite{quench6}.

\section*{Acknowledgments}

We thank N. V. Vysotina for her valuable help with numerical simulations.
The work of B.A.M. is supported, in a part, by grant No. 2015616 from the
joint program in physics between the National Science Foundation (US) and
Binational Science Foundation (US-Israel). This author appreciates
discussions with R. G. Hulet, M. Olshanii, and V. A. Yurovsky.

\end{document}